# Flying Particle Microlaser and Temperature Sensor in Hollow-Core Photonic Crystal Fiber

RICHARD ZELTNER*, RICCARDO PENNETTA, SHANGRAN XIE, PHILIP ST.J. RUSSELL

*Max Planck Institute for the Science of Light, Staudtst.2, 91058 Erlangen*
*Corresponding author: richard.zeltner@mpl.mpg.de*

Whispering-gallery mode (WGM) resonators combine small optical mode volumes with narrow resonance linewidths, making them exciting platforms for a variety of applications. Here we report a flying WGM microlaser, realized by optically trapping a dye-doped microparticle within a liquid-filled hollow-core photonic crystal fiber (HC-PCF) using a CW laser and then pumping it with a pulsed excitation laser whose wavelength matches the absorption band of the dye. The laser emits into core-guided modes that can be detected at the endfaces of the HC-PCF. Using radiation forces, the microlaser can be freely propelled along the HC-PCF over multi-cm distances—orders of magnitude further than in previous experiments where tweezers and fiber traps were used. The system can be used to measure temperature with high spatial resolution, by exploiting the temperature-dependent frequency shift of the lasing modes, and also for precise delivery of light to remote locations.

## 1. INTRODUCTION

Whispering-gallery mode (WGM) resonators are powerful tools for a variety of applications including quantum optics, nonlinear optics [1] and sensing [2]. By exploiting the sensitivity of the cavity resonances to changes in the resonator geometry and environment, sensing of physical quantities [3,4] as well as chemical and biological samples even down to the single molecule level [5], has been demonstrated. In most experiments passive resonators are used, relying on evanescent excitation of WGMs using tapered waveguides or prism coupling [6]. These techniques require close proximity between the resonator and the coupler. In contrast, active WGM resonators containing a gain medium permit remote excitation and collection of the emitted spectrum, even if the resonator is placed in inaccessible environments such as biological samples [7]. Moreover, when operated above threshold, linewidth narrowing of the resonances and increased oscillator strength improves the sensing performance compared to passive resonators [8].

Optical tweezers and related techniques [9,10] allow precise control of the position of microparticles, potentially permitting position-dependent WGM-based sensing [11]. It was shown that active WGM resonators can lase when captured in an optical tweezer or fiber trap [12,13]. Although considerable effort has been made to enhance the manipulation range of optical tweezers and fiber traps, it is limited to a few mm or less by the Rayleigh range or the dimensions of the optical system, even when complex experimental configurations involving spatial light modulators [14] or non-diffracting Bessel beams are used [15]. Hollow-core photonic crystal fiber (HC-PCF) allows optical trapping and propulsion of individual microparticles within the fiber core over distances limited only by the fiber loss. The particle can be freely moved along the fiber axis and is protected from unwanted external perturbations [16]. Recently, a "flying particle sensor" using trapped particles inside HC-PCF was reported [17,18]

Here we report a "flying" WGM microlaser consisting of a dye-doped particle optically trapped and propelled within the core of a liquid-filled HC-PCF. The microparticle laser is pumped by sub-ns pulses at 532 nm, which are launched into the fundamental core mode along with a CW trapping beam at 1064 nm. Laser emission at ~590 nm is collected at the fiber endfaces. Using the thermally-induced frequency-shift of the lasing modes, we show that this flying WGM microlaser can be used to remotely measure the temperature distribution along a HC-PCF with a spatial resolution of ~5 mm.

## 2. EXPERIMENTAL SETUP

The experimental setup is sketched in Fig. 1. One end of a kagomé-style HC-PCF with 30 μm core diameter (SEM shown in the inset) was mounted inside a liquid cell connected to a syringe pump to allow filling with $D_2O$. The typical fiber length in the experiment was ~30 cm. The loose end was placed inside a polydimethylsiloxane (PDMS) chamber containing a solution of commercial dye-doped polystyrene microparticles ($\lambda_{ex}$ = 532 nm, $\lambda_{em}$ = 570 nm from microParticles GmbH) with diameter of ~ 15.5 μm. Light was launched into the HC-PCF via glass windows on the PDMS chamber and the liquid cell, and the system was optimized to ensure that mainly the $LP_{01}$-like core mode was excited.

An external optical tweezer system (operating at 1064 nm wavelength, not shown in Fig. 1) was installed below the PDMS chamber and used to trap a single particle and position it in front of the fiber core. Once the particle was trapped, the tweezing power was reduced and the 1064 nm propulsive power increased, propelling the particle into the HC-PCF core and trapping it at core center. The launched trapping beam power was typically ~500 mW. At the same time 532 nm pump pulses (800 Hz repetition rate) were launched in the reverse direction into the fiber. A portion of the emitted laser light at ~590 nm was captured by backward-propagating core modes and guided towards the fiber end, where it was separated from the pump and trapping light and analyzed using a spectrometer (spectral resolution 0.5 nm). The speed and position of the lasing particle were monitored using both Doppler-velocimetry [17] and an imaging system mounted above the HC-PCF. The typical particle speeds in the experiment were between 100 and 250 μm/s. The particle could be held stationary at any position along the fiber by introducing a liquid counter-flow [16] or by launching a second 1064 nm laser beam into the counter-propagating core mode.

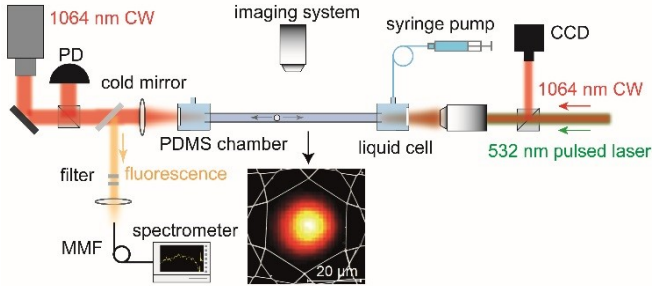

Fig. 1 Experimental setup. PD, photodiode; MMF, multimode fiber. The inset shows the measured near-field image of the fiber modes at the endface of the HC-PCF at the trapping wavelength (1064 nm) superimposed on a scanning electron micrograph (SEM) of HC-PCF structure.

## 3. LASING PROPERTIES

Fig. 2(a) shows the measured optical emission spectra of a single particle trapped stationary approximately 6 cm from the fiber entrance by a liquid counter-flow. With increasing pump power three peaks emerge at wavelengths $\lambda_1 = 589.7$ nm, $\lambda_2 = 594.9$ nm and $\lambda_3 = 600.4$ nm. These wavelengths correspond to WGMs with angular orders 117 to 119 and radial order 1, or angular orders 110 to 112 and radial order 2 [19]. Lorentzian fits to the measured peaks yield linewidths around 3.5 nm, corresponding to a Q-factor of ~175. This value is most likely to be limited by the spectral resolution of the spectrometer. Fig. 2(b) plots the integrated spectral counts for each peak (proportional to energy below each peak) as a function of launched average pump power, revealing a clear threshold at ~8 μW (corresponding to ~5 W peak power at the particle) for all three peaks. We attribute the varying slope efficiencies to the spectral dependence of the gain. Note that for the integration of the spectral counts, for each individual pump power, the collected emission spectra of the particle were corrected by subtracting a reference spectrum measured without a particle, so as to remove the influence of effects such as Raman scattering in the $D_2O$.

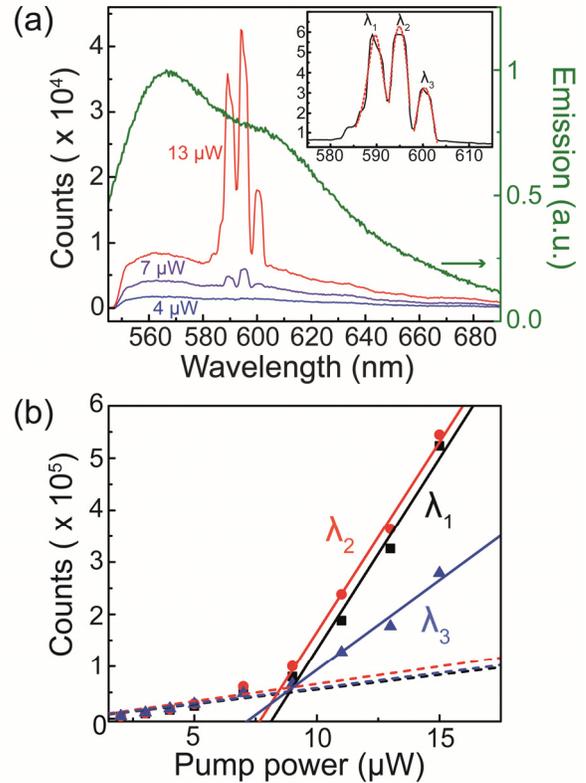

Fig. 2 (a) Measured emission spectra with increasing average pump power for a dye-doped polystyrene microparticle trapped in a HC-PCF. The dotted green curve shows the fluorescence spectrum of the particle. The inset shows the zoom-in of the lasing peaks and its Lorentzian fit (red-dashed curve). (b) Integrated spectrometer counts for each peak as a function of average pump power, showing a lasing threshold of ~8 μW.

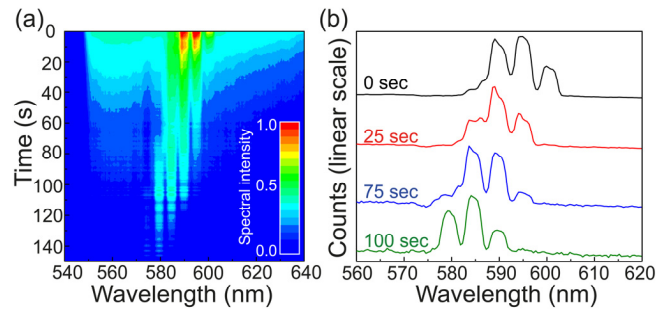

Fig. 3 (a) Temporal evolution of the emission spectrum from the particle. (b) Normalized emission spectra at four different times under irradiation. The decrease in amplitude of the lasing peaks, accompanied by a spectral blue-shift, is a fingerprint of dye lasing.

To confirm the presence of dye-related gain, the particle was held stationary and pumped continuously well above threshold. The spectral evolution over time (Fig. 3(a)) shows a decrease in the amplitude of the lasing peaks on a timescale of tens of seconds, caused by photobleaching of the dye molecules. The measured spectra at four instants (Fig. 3(b)) show that the shorter-wavelength lasing modes become stronger during the bleaching process, which reduces the absorption at shorter wavelengths [12].

Optical scattering forces created by the 1064 nm trapping beam permit the particle to be translated along the fiber. The lasing spectra at four different axial positions are plotted in Fig. 4(a), and video frames of the 1064 nm light scattered by the particle, taken through the side of the fiber, are shown in

Fig. 4(b). At each position the spectrum was obtained with the particle held stationary using a liquid counter-flow. While the particle was being moved, the 532 nm pump laser was blocked to avoid photobleaching. The measurements show that the lasing wavelengths are independent of particle position. Notably, the lasing peak at ~597 nm shows a split with a separation of ~2.6 nm, which matches quite well the predicted separation between TE and TM modes with angular mode numbers 117 to 119 and radial mode number 1 (~2.3 nm).

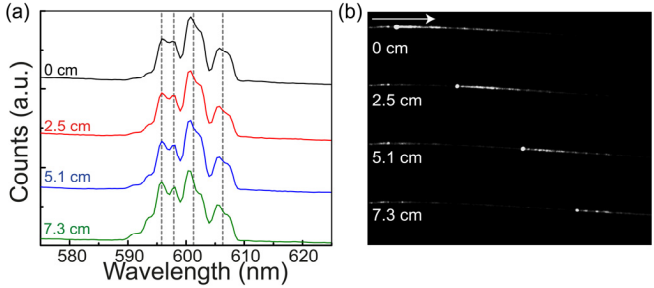

Fig. 4 (a) Lasing spectra measured at different axial positions of the particle. (b) Individual video frames collected through the side of the HC-PCF, showing the lasing particle at different positions. The particle is visible due to scattered 1064 nm trapping light. The white arrow indicates the propagation direction of the trapping laser beam.

## 4. REMOTE THERMAL SENSING

The results in Fig. 4 suggest that the flying microlaser might be used as a WGM-based sensor, for example for temperature. To this end, a section of the HC-PCF was placed in a metal V-groove connected to a heat source. A dye-doped microparticle was moved into and trapped motionless within the heated region using a counter-propagating trapping beam. In this experiment, we used dye-doped melamine-resin particles (microParticles GmbH) with a diameter of ~13.5 μm. Their absorption, emission and lasing properties (see SM1) were similar to those of the polystyrene particles described above, although their smaller diameter and higher refractive index meant that they could be trapped and propelled at lower optical power, reducing the effects of heating caused by absorption of trapping light in the $D_2O$ [20] and the particle.

The lasing spectra at three different temperatures show a clear blue-shift in the emission wavelength (Fig. 5(a)), caused by the negative thermo-optic coefficient of the WGMs, which dominates the temperature response [4]. Fig. 5(b) plots the wavelength shift with increasing temperature of the mode that initially lases at ~610.2 nm at 25°C. The response is linear, with a slope of –0.122 nm/K. Different lasing modes and particles were found to show comparable sensitivities. The detection limit of the sensor, obtained by dividing the resolution by the sensitivity [21], works out at ~2.7 K, where we estimated the resolution as three times the average standard deviation of all the measurements.

Next, a second heater was placed ~18 mm from the first, and the temperatures of both set to ~22 K above room temperature. Fig. 5(c) plots the wavelength shift as a function of distance (left axis), monitored while propelling the particle over both heaters while continuously recording the lasing spectra. Two distinct regions with higher temperature can be identified, separated by a distance that agrees well with the separation of the heaters. Using the calibration in Fig. 5(b), the temperature profile could be reconstructed (right axis) with a spatial resolution of ~5 mm, showing a maximum temperature increase of ~25 K. The red curve in Fig. 5(c) shows the temperature distribution at thermal equilibrium, simulated by finite element modelling (see SM2). The first data-point of the measurement was used as a reference, and the agreement is good. We attribute the oscillations in the resonance shift to the presence of higher order modes in the trapping light, which cause intermodal beating that periodically modulates the particle temperature (see SM3).

## 5. CONCLUSIONS

Dye-doped WGM microlasers can be optically trapped and propelled inside a hollow-core photonic crystal fiber and used for distributed sensing of temperature over cm length-scales with a spatial resolution of a few mm. Photobleaching of the dye, which limits the current performance, could be avoided using inorganic gain media such as quantum dots or rare earth ions [22]. The unique combination of particle guidance in HC-PCF and WGM-based sensing extends the range of potential applications of active WGM systems, specifically opening up the possibility of remote probing of physical quantities with high spatial resolution and sensitivity. The flying WGM microlasers may also be useful for spatially precise delivery of light to remote locations.

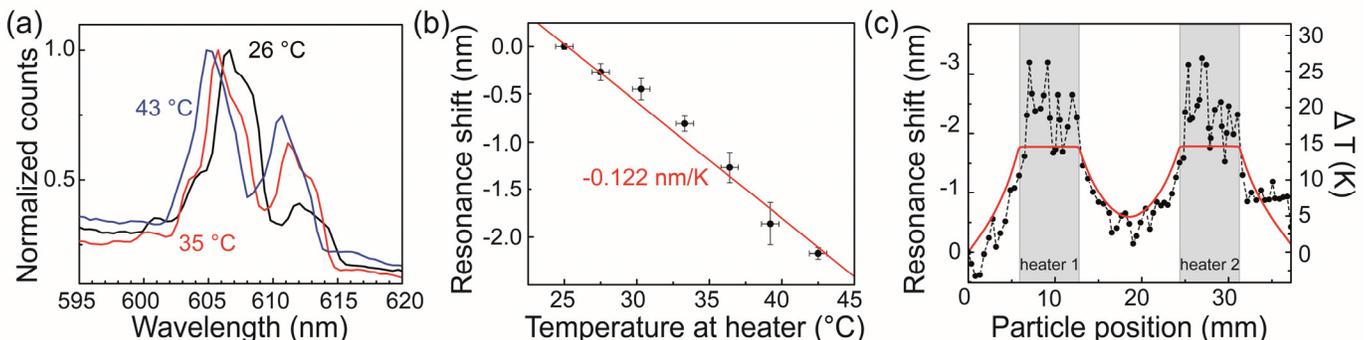

Fig. 5 (a) Lasing spectra of an optically trapped dye-doped melamine-resin particle at different temperatures, showing a blue-shift with increasing temperature. (b) Measured wavelength shift of the 610 nm lasing mode as a function of temperature, with linear fit (red). (c) Wavelength shift of the lasing mode (left axis) and reconstructed temperature profile (right axis) along the fiber when the micro-laser was moved over two heaters set to ~22 K above room temperature. The red curve plots the simulated temperature distribution using finite element modeling.

# Supplementary material

## S1. Lasing spectra of melamine-resin particles

To characterize the lasing properties of the melamine-resin particles, a single particle was launched into the core of HC-PCF and was hold stationary approximately 6 cm from the fiber entrance. Fig. S1 plots the measured emission spectrum of a single melamine-resin inside HC-PCF. For increasing pump power three peaks are emerging at wavelengths $\lambda_1 = 607.1$ nm, $\lambda_2 = 612.5$ nm, and $\lambda_3 = 618.0$ nm.

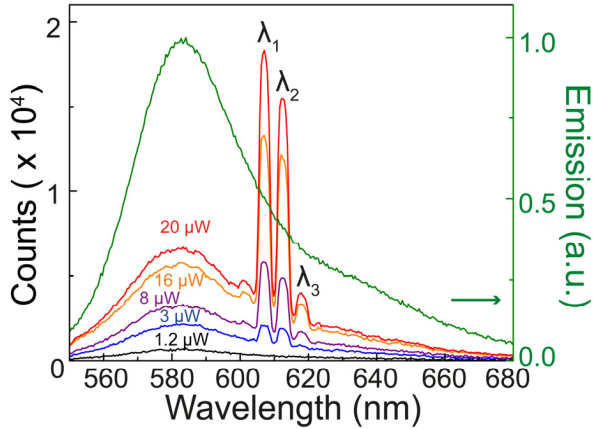

Fig. S1 Emission spectra of an individual melamine-resin particle trapped stationary inside the HC-PCF. The green curve plots the emission spectrum of a bulk solution of the melamine-resin particles.

## S2. Finite element modelling of the temperature distribution along HC-PCF

When heaters were placed along the HC-PCF, the temperature distribution was modelled using the heat transfer module of COMSOL Multiphysics. The HC-PCF was modelled as a fused silica capillary with the same outer diameter as the fiber we used, with the inner diameter set to the diameter of the photonic structure. The hollow region of the capillary was filled with $H_2O$. The two heaters were aluminum parts with the same dimensions as used in the experiments. The heaters were placed on an iron block with heat source boundary condition. The heat source was set such that the temperature on the surface of the heaters was matching the temperature measured with the thermometer. To account for air-flow in the lab during the experiment we assumed a normal laminar inflow and outflow of 15 cm/s, cooling the free hanging part of the HC-PCF between the heaters. The heaters were surrounded by an airbox set at the temperature which was measured at the heaters when no heating was applied. To justify the value of the air-flow, we compared the simulation results with an experiment during which a lasing particle was moved over only one heater. Fig. S2 plots the experimentally reconstructed temperature profile (black dots) together with the simulation (red curve) using one heater. They agree well with each other.

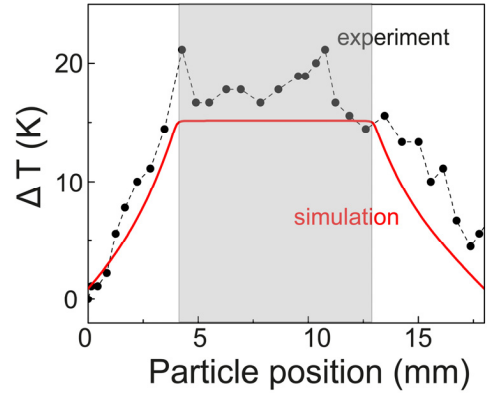

Fig. S2 Simulated temperature profile (red curve), compared with the experimental result (black dots) in which the particle was moved over a single heater.

## S3. Analysis of the oscillations in resonance shift

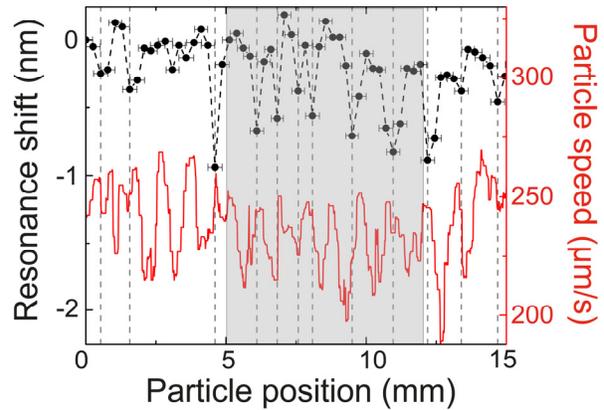

Fig. S3 Resonance shift as a function of particle position when a lasing particle was moved over the heater (indicated with the grey area) set at RT. The red curve plots the particle speed, revealing speed oscillations caused by intermodal beating of the fiber modes at trapping wavelength.

To investigate the origin of the observed oscillations in the resonance shift, a lasing particle was moved over one heater at room temperature. Fig. S3 plots the measured resonance shift as a function of particle positon (black dots). Clear oscillations are again visible, which become more pronounced in the area where the particle is passing the heater (grey-shaded area). The red curve plots the particle's speed, which also shows oscillations caused by a periodically varying optical intensity along the axis of the HC-PCF. In regions of high intensity the particle experiences higher optical scattering forces and consequently speeds ups. The oscillation period in the particle speed equals $L_B/2$ where $L_B$ is the beat length between the waveguide modes [1]. Strikingly, the position of maxima in the resonance shift coincide fairly well with maxima of particle speed. The corresponding periods of ~1.2 mm in both measurements are in good agreement with the predicted half-beat length $L_B/2$ of 1.25 mm between the fundamental mode ($LP_{01}$) and the first higher order mode ($LP_{11}$) [2] which are typically most efficiently excited in the experiment. We thus conclude that the oscillations are an artefact caused by periodic heating of the particle when it is moving through the intermodal beat pattern of the trapping laser. The optically propelled particle scatters a considerable fraction of light (experimentally measured value ~15%) into the surrounding of the HC-PCF [3]. When the particle is passing the heater a part of the scattered light is reflected by the metallic surface and may be re-absorbed by the particle and $D_2O$, explaining why the oscillations are more pronounced when the particle is passing the heater.